# Nuclear forensics using gamma-ray spectroscopy

Eric B. Norman

*Department of Nuclear Engineering, University of California, Berkeley, CA 94720 U. S. A.*

**Abstract.** Much of George Dracoulis's research career was devoted to utilizing gamma-ray spectroscopy in fundamental studies in nuclear physics. This same technology is useful in a wide range of applications in the area of nuclear forensics. Over the last several years, our research group has made use of both high- and low- resolution gamma ray spectrometers to: identify the first sample of plutonium large enough to be weighed; determine the yield of the Trinity nuclear explosion; measure fission fragment yields as a function of target nucleus and neutron energy; and observe fallout in the U. S. from the Fukushima nuclear reactor accident.

## 1 Seaborg's Plutonium?

In early 1941, in Room 307 Gilman Hall on the University of California's Berkeley campus, Glenn Seaborg and his collaborators Arthur Wahl and Joseph Kennedy chemically separated element 94 from uranium that had been bombarded with 16-MeV deuterons at Berkeley's 60" cyclotron. Not long after, Seaborg's group proposed the name plutonium for this new element. For the next year and a half, Seaborg's group performed numerous studies of both the nuclear and chemical properties of plutonium using only trace amounts of this new material. In order to study plutonium in its pure form, an effort was therefore made to produce a "macroscopic" amount of plutonium by irradiating many kilograms of natural uranium with neutrons produced by bombarding beryllium with accelerated deuterons. Cyclotrons at both Washington University and Berkeley were utilized for this production. After chemically extracting the plutonium from the uranium and fission products, on September 10, 1942, at the Metallurgical Laboratory of the University of Chicago, Boris Cunningham and Lewis Werner succeeded in producing the first sample of pure $PuO_2$ that was large enough to be weighed on a newly developed microbalance. The mass of the $PuO_2$ was determined to be 2.77 μg. This sample was preserved by sealing the platinum boat and oxide deposit inside a glass tube. The subsequent history of this sample is not known, but it was on display for a number of years at the Lawrence Hall of Science in Berkeley, CA.

At some point in the late 2000's, as a result of both financial and perceived radiation safety concerns, the Lawrence Hall staff decided to remove this sample from display. In 2008, a plastic box with a label on it stating "First sample of Pu weighed. 2.7 μg" was found at the Hazardous Material Facility on the University of Berkeley's campus. It was assigned an EH&S sample number S338. There it was could have been discarded as radioactive waste were it not for the knowledgeable eye of Phil Broughton, from UC Berkeley's Environmental Health and Safety Department. The paper trail documenting this sample's history had been lost and so the question was what could be done to establish its authenticity as Seaborg's plutonium?

In an attempt to authenticate this sample without opening the plastic box, we performed passive x-ray and gamma–ray analysis using both planar and coaxial high-resolution Ge detectors. A background-subtracted gamma-ray spectrum observed in approximately one day of counting is shown in Figure 1. The object was found to contain $^{239}$Pu. No other radioactive isotopes were observed. We carefully measured our gamma-ray detection efficiency for this sample and thus were able to quantitatively determine the mass of $^{239}$Pu contained in the box to be $2.0 \pm 0.3$ μg. These observations are consistent with the identification of this object being the 2.77-μg $PuO_2$ sample described by Glenn Seaborg and his collaborators as the first sample of $^{239}$Pu that was large enough to be weighed [1].

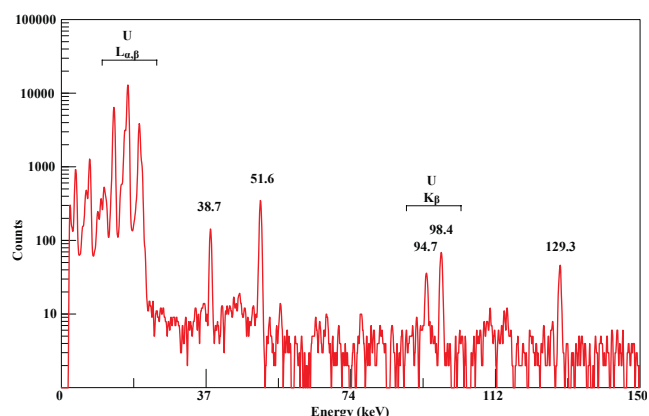

Figure 1. Background subtracted spectrum observed in our planar Ge detector from Sample S338. All of the labeled peaks are x-rays and gamma rays attributable to the decay of $^{239}$Pu.

## 2 Yield of the Trinity nuclear explosion

In 1945, the first nuclear explosive device was tested in Alamogordo, New Mexico. It was an implosion-type device using $^{239}$Pu as the fissile material. Following the explosion, fission products and other materials contained in the "gadget" along with dirt and debris from the surrounding desert were caught up in the resulting fireball and heated to extremely high temperatures. When this material cooled and fell to the ground some of it was incorporated into a greenish glassy material now called Trinitite. We obtained four samples of Trinitite from Mineralogical Research Co. in San Jose, CA. These samples ranged in mass from 9 to 16 grams and were each counted using a planar Ge detector. Portions of the spectra observed from one of these samples are shown in Figures 2 and 3.

---

[a] Corresponding author: ebnorman@lbl.gov

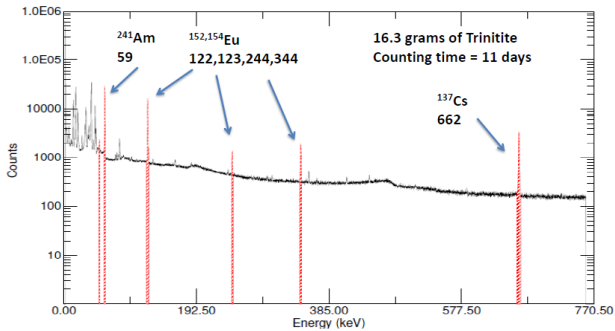

Figure 2. Energy spectrum observed from the 16.3-gram Trinitite sample in 11 days of counting. The 662-keV line from the decay of the long-lived fission product $^{137}$Cs is clearly seen.

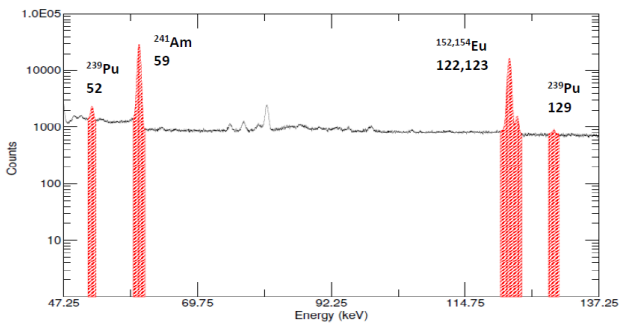

Figure 3. Low-energy portion of the spectrum observed from the 16.3-gram Trinitite sample. The 59- and 129-keV lines from the decay of $^{239}$Pu are clearly seen.

As can be seen in these figures, we observed the characteristic $^{239}$Pu gamma-ray lines at 59 and 129 keV along with the 662-keV line from the long-lived fission product $^{137}$Cs. All other fission products would have decayed to undetectable levels (except for $^{90}$Sr that does not emit any gamma rays). We carefully measured our gamma-ray detection efficiencies for the extended and unusual-shaped samples and then determined the activities of $^{239}$Pu and $^{137}$Cs contained in each one. Based upon these results, we then performed a very simple-minded analysis to determine the explosive yield of the Trinity device. To do so, we assumed the same probability for Cs and Pu to become incorporated in Trinitite. Due to high volatility of Cs, this should lead us to underestimate the yield. We also neglected the fission yield from the tamper of depleted uranium. This would also cause us to underestimate the yield. There is a very interesting interplay of both physics and chemistry involved in this analysis. The results from our four samples produce yields that range from 3.5 – 9.8 kilotons. These are all substantially lower than the actual yield of 21 kilotons, as could be expected from the type of analysis we carried out, but are at least the correct order of magnitude.

## 3 Beta-delayed gammas from fission

When a nucleus undergoes fission, the products emit both prompt and delayed characteristic gamma-rays. The delayed gamma-rays are emitted following beta-decays, and therefore also have characteristic time scales associated with their emissions. These gamma-rays can be used to identify fissionable material hidden in cargo containers, or distinguish between fissionable isotopes, and are useful for assay of nuclear fuels. They can also be potentially used in nuclear forensics to identify the average energy of neutrons producing fission and the original fissionable isotope.

However, for beta-delayed gamma-ray spectroscopy to be a useful tool, a complete set of measurements on all likely fissionable isotopes is needed. We, therefore, performed measurements of beta-delayed gammas following the thermal and 14-MeV neutron-induced fissions of $^{235}$U and $^{239}$Pu and also the fast neutron induced fissions of $^{232}$Th, $^{237}$Np and $^{238}$U. We used both high-resolution germanium detectors and low-resolution plastic scintillators to measure gamma energy and time spectra. We found distinctive gamma-ray intensity ratios for each sample in several time bins, which provide adequate data to identify the fissionable material present and can also provide information on the neutron energy [2, 3].

Figure 4 illustrates some of the differences we observed following the thermal-neutron induced fissions of $^{235}$U (shown on top) and $^{239}$Pu (shown below). Similar differences in gamma-ray line intensities were observed at a number of different energies form 5 minutes to 14 hours after fission. Using ratios of gamma-ray line intensities, we showed that one can readily determine which isotope underwent fission. Similarly, one can use gamma-ray line intensity ratios to learn about the energy of the neutron that induced the fission. Figure 5 illustrates some of the differences we observed following the thermal and the 14-Mev neutron induced fission of $^{235}$U.

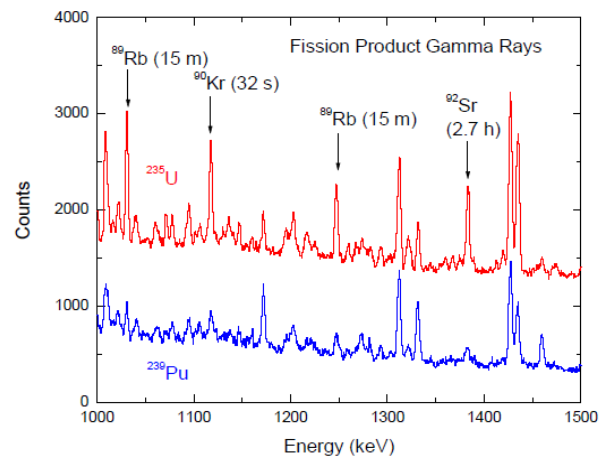

Figure 4. Arrows indicate lines from isotopes produced more strongly in $^{235}$U fission than $^{239}$Pu fission.

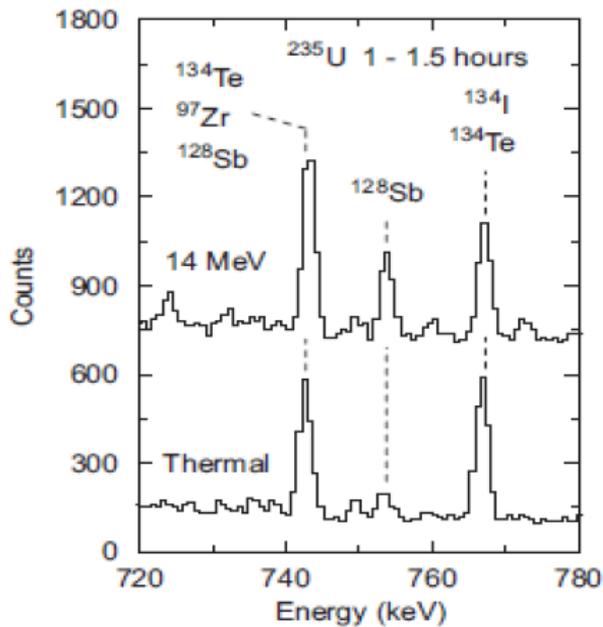

Figure 5.  Thermal versus 14-MeV fission of $^{235}$U at 1 – 1.5 hours after fission.

## 4 Fukushima fallout in the US

In order to search for possible worldwide distribution of radioactive contamination following the accident at the Fukushima Dai-ichi nuclear power plant in Japan, we collected rainwater samples in Berkeley, Oakland, and Albany, California and examined them for the presence of above normal amounts of radioactivity.  Samples were collected from March 16 – March 26, 2011, by placing plastic containers outside in the Oakland and Berkeley hills and in Albany. Collection periods varied from a few to approximately 12 hours.  Following collection, each sample was placed directly into a Marinelli beaker for gamma-ray counting.  No chemical or physical processing of any kind was done to the rainwater samples before counting.

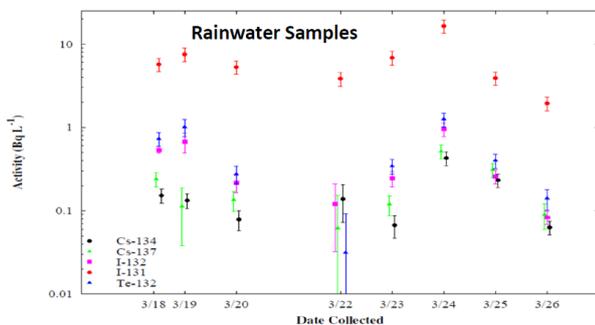

Figure 6.  Activities of $^{131,132}$I, $^{132}$Te, and $^{134,137}$Cs in Bq/liter measured in San Francisco Bay area rain water as a function of time.

The first sample which showed activity above background was collected on March 18, 2011.  Our spectra indicated the presence of $^{131,132}$I, $^{132}$Te, and $^{134,137}$Cs.  These are relatively volatile fission fragments produced with large cumulative yields from the fission of $^{235}$U and $^{239}$Pu.  The levels we observed were low and were not a health risk to the public [4]. Subsequently we collected and counted numerous food samples of fish, seaweed, and dairy products being sold in the San Francisco Bay Area.  Portions of the spectra we observed from samples of weeds collected in April, 2011 in Oakland, CA and from Philippine tuna fish sold in Berkeley, CA in 2013 are shown in Figure 7.  We did not observe any evidence of Fukushima contamination in any of the samples studied.  We did, however, observe very low levels of $^{137}$Cs in a number of samples that probably originated from atmospheric nuclear weapons tests conducted in the 1950's and 1960's [5].

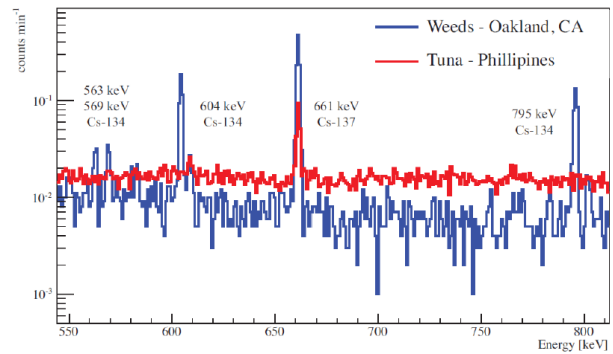

Figure 7.  Comparison of gamma spectra of local weeds collected in April 2011 in Oakland, CA along with a sample of Tuna from the Philippines purchased locally at a San Francisco Bay Area retail location.

## Conclusions

The measurements discussed in this paper illustrate how, in addition to its utility in fundamental studies in nuclear physics,  gamma-ray spectroscopy can be used to study a number of interesting problems in nuclear forensics.

## Acknowledgements

I wish to thank all of my collaborators in these projects.  This work was supported in part by the U. S. Department of Energy National Nuclear Security Administration under Award Number DE-NA0000979 and by the U. S. Department of Homeland Security.

## References

[1]  E. B. Norman, K. J. Thomas, K. E. Telhami, Amer. Journ. Phys. **83**, 843 (2015)
[2]  R. E. Marrs *et al*., Nucl. Instrum. Meth. A **592**, 463 (2008)
[3]  A. Iyengar *et al.,* Nucl. Instrum. Meth. B **304**, 11 (2013)
[4]  E. B. Norman, C. T. Angell, P.  A. Chodash, PLoSONE  **6**(9):e24330 (2011)
[5]  A. R. Smith *et al.,* Journ. of Environmental Protection **5**, 207 (2014)